# Intrinsic gaps in the classical theory of p-n junctions

Jianhong Yang,[1,*] Tinglan Li,[1] and Gonghao Sun[1]

[1]*School of Physical Science and Technology, Lanzhou University, Lanzhou 730000, China*

*Contact author: yangjh@lzu.edu.cn

The classical analytical theory of p–n junctions successfully describes minority-carrier injection and predicts the ideal diode characteristic, but its conventional reduced formulation leaves the connection between internal transport and terminal response incompletely resolved. This paper critically examines five interrelated gaps in that formulation: the electrostatic justification of charge quasi-neutrality, the determination of the quasi-neutral-region electric field, the explicit treatment of majority-carrier transport, the spatial reconciliation of electron and hole currents, and the evaluation of terminal current from the resolved local transport. A review of representative foundational, textbook, and research treatments distinguishes recognition of individual issues from their integration into a self-consistent analytical description. The central finding is that these gaps constitute a common analytical closure problem: carrier densities, the internal field, and current partition must be determined consistently rather than treated as separate elements. This diagnosis concerns the reduced analytical description, not the validity of the underlying drift–diffusion equations or charge conservation. A two-carrier framework is outlined in which electrostatic consistency, coupled transport, and junction-edge and contact conditions connect local current balance to terminal response. The objective is to preserve the classical diode equation within its established domain while making its physical basis and approximation limits explicit.

## I. INTRODUCTION

### A. Historical context and the classical framework

The p-n junction is the basic active element in diodes, bipolar transistors, solar cells and laser diodes. In each case the terminal characteristics follow from carrier injection and transport across one metallurgical junction between p- and n-type material.

The classical theoretical framework for this device was developed in the years immediately following the invention of the transistor. Shockley's 1949 paper [1] established the foundational formulation that has since become universal: the injection and diffusion of minority carriers across the junction, the Shockley boundary conditions relating minority-carrier densities at the depletion-region edges to the applied voltage, and the derivation of the ideal diode equation $J = J_0\left(e^{qV/kT} - 1\right)$. The elegance, simplicity, and predictive power of this framework secured its immediate acceptance, and it has been reproduced, refined, and elaborated in every major semiconductor textbook ever since [2-5].

Extensions and refinements followed rapidly. Sah, Noyce, and Shockley [6] incorporated generation-recombination currents within the space-charge region, explaining the ideality factor in real diodes. Hall [7] and Shockley and Read [8] established the statistics of carrier recombination through traps, providing the microscopic basis for junction recombination currents. Moll [9] and Ebers and Moll [10] extended the framework to transistor operation and developed the charge-control model. These contributions collectively constitute what is now known as the standard, or classical, theory of p-n junctions.

The classical theory remains useful, but its reduced formulation leaves several structural gaps within the low-injection, long-base framework itself. These are not peripheral inaccuracies arising from neglected second-order effects (high-level injection, heavy doping, or surface recombination) that have been the subject of an extensive research literature [11-14]. They are, rather, fundamental gaps within the low-level-injection, long-base framework that constitutes the core of the classical theory itself. It does not say how charge quasi-neutrality is established, what field exists in the base, how majority carriers carry current, how continuity is maintained, or whether the junction voltage $V_j$ is necessarily a part of the applied bias $V_a$.

## B. Scope and objectives of the present work

This paper has a threefold objective. First, to provide a precise, systematic identification and characterization of the five intrinsic gaps of the classical theory. Second, to survey the existing literature and assess the degree to which each gap has, or has not, been addressed. Third, to motivate the construction of a complete self-consistent analytical theory resolving the gaps simultaneously.

The paper is organized as follows. Sec. II reviews the mathematical structure of the classical theory to establish notation and to locate precisely where the incompleteness resides. Sec. III presents a detailed analysis of each of the five gaps. Sec. IV surveys the literature. Sec. V looks ahead to possible solutions. Sec. VI concludes.

## II. THE CLASSICAL DRIFT-DIFFUSION FRAMEWORK

The drift–diffusion–Poisson system comprises the electron and hole transport equations, carrier continuity equations, and Poisson's equation, supplemented by appropriate boundary conditions. The classical analytical description introduces additional approximations to simplify this coupled system. The limitations examined here concern the consistency and scope of these approximations, not the underlying governing equations.

Consider a one-dimensional, abrupt (step) p-n junction, with the metallurgical junction at $x = 0$, as shown in Fig. 1. The depletion, or space-charge, region (SCR) spans $[-x_{p0}, x_{n0}]$; the quasi-neutral p-region (QNR-p) occupies $[x_{p0}, W_p]$, and the quasi-neutral n-region (QNR-n) occupies $[x_{n0}, W_n]$. Uniform doping is assumed: acceptor concentration $N_\mathrm{A}$ in the p-region and donor concentration $N_\mathrm{D}$ in the n-region, giving equilibrium minority-carrier concentrations $n_{p0} = n_i^2/N_\mathrm{A}$ and $p_{n0} = n_i^2/N_\mathrm{D}$.

The complete drift-diffusion description of carrier transport requires three coupled sets of equations. The transport equations are

$$J_n(x) = q\mu_n n(x)\,E(x) + qD_n\frac{dn}{dx}, \tag{1}$$

$$J_p(x) = q\mu_p p(x)\,E(x) - qD_p\frac{dp}{dx}. \tag{2}$$

where $E(x)$ is the electric field, $\mu_n$ and $\mu_p$ are the carrier mobilities, and $D_{n,p} = (kT/q)\mu_{n,p}$ are the Einstein-relation diffusivities. The steady-state continuity equations are

$$\frac{1}{q}\frac{dJ_n}{dx} = R - G, \qquad -\frac{1}{q}\frac{dJ_p}{dx} = R - G. \tag{3}$$

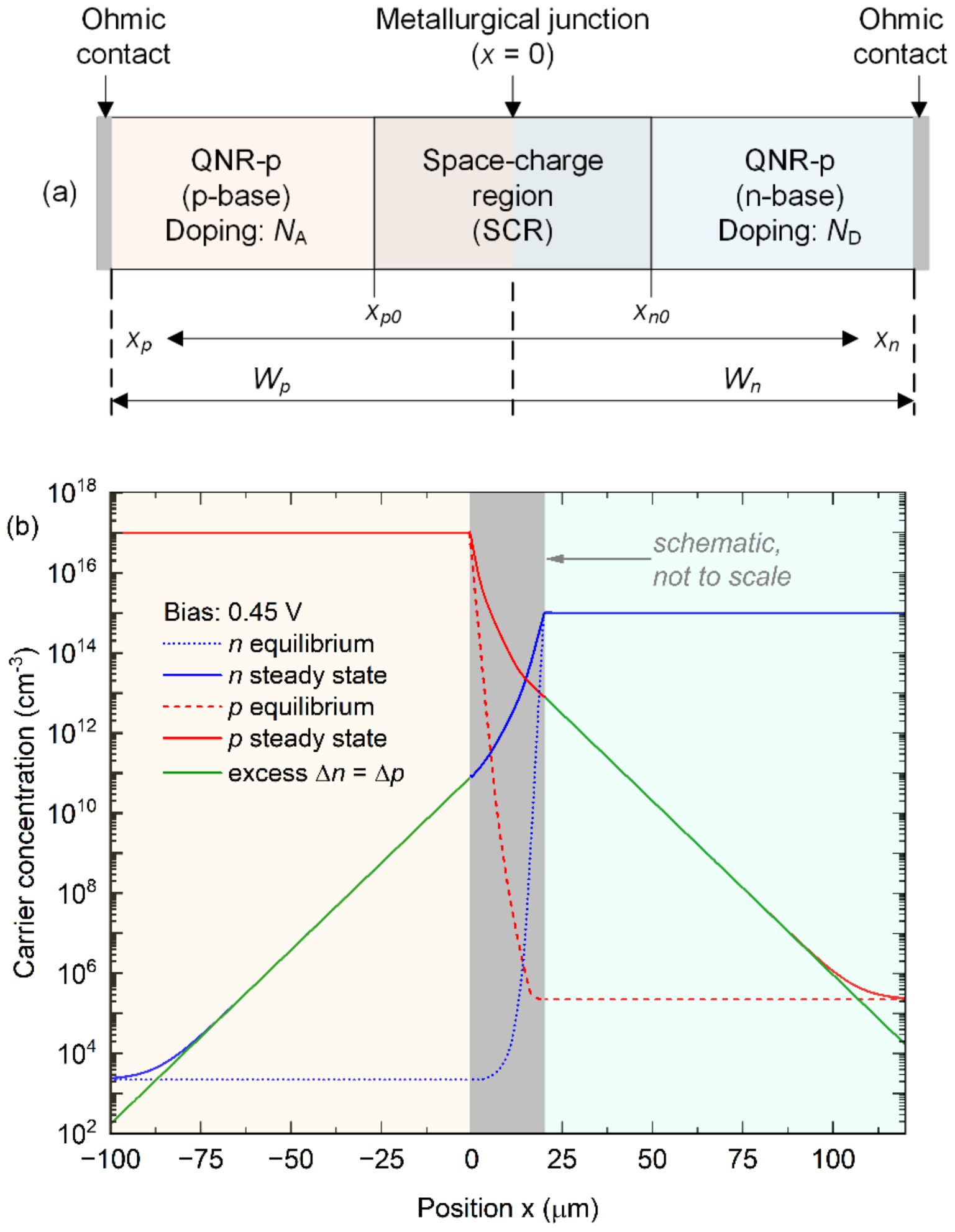

**FIG. 1.** Reference geometry and carrier distribution of the one-dimensional abrupt p-n junction used throughout this paper. (a) Physical structure and coordinate system: the metallurgical junction is at $x = 0$; the space-charge region (SCR) spans $[-x_{p0}, x_{n0}]$; the QNR-p and QNR-n extend to the ohmic contacts at $x = -x_p$ and $x = x_n$ respectively. (b) An example of steady-state carrier distributions in a junction (Si, 300 K; $N_A = 10^{17}$ cm$^{-3}$, $N_D = 10^{15}$ cm$^{-3}$; base widths $W_p = W_n = 200\ \mu$m) under bias 0.45 V. The built-in voltage is $V_{bi} = 0.693$ V. The SCR width is smaller than 0.95 μm but is schematically exaggerated for illustration. The carrier steady-state concentrations are the sum of equilibrium and excess concentrations for each species of carriers.

where $R$ and $G$ are the recombination and generation rates. Poisson's equation closes the system:

$$\frac{dE}{dx} = \frac{q}{\varepsilon_s}(p - n + N_D - N_A). \tag{4}$$

The total current density $J = J_n + J_p$ is spatially uniform in steady state as a direct consequence of Eq. (3), which is the mathematical statement of current continuity.

The classical theory applies distinct approximations by region. In the depletion region, the depletion approximation (neglect of mobile carriers) reduces Poisson's equation to a piecewise-linear field profile, yielding the built-in voltage $V_{bi}$, the depletion widths, and the junction capacitance [2, 3, 15]. In the quasi-neutral regions (QNRs), three key simplifications are imposed: (i) charge neutrality, $p - n + N_\mathrm{D} - N_\mathrm{A} = 0$, is assumed, implying $dE/dx = 0$ via Poisson's equation; (ii) the QNR electric field is set to zero; (iii) only the minority-carrier transport equation is solved, yielding the diffusion equation for the excess minority holes in QNR-n (with $\Delta p_n \equiv p_n - p_{n0}$) for example:

$$D_p \frac{d^2 \Delta p_n}{dx^2} - \frac{\Delta p_n}{\tau_p} = 0, \tag{5}$$

subject to Shockley's boundary condition $\Delta p_n(x_{n0}) = p_{n0}\left(e^{qV/kT} - 1\right)$, where $V$ is the voltage drop across the junction itself. The resulting exponential minority-carrier profile, evaluated at the depletion-region edge, gives the hole-current contribution. An analogous procedure in QNR-p gives the electron contribution. The sum of hole and electron contributions is the Shockley diode equation. Fig. 2 displays these classical excess-carrier profiles and the associated diffusion currents. The classical framework is thus an exclusively minority-carrier theory: it treats minority-carrier injection and diffusion, while majority-carrier transport is never evaluated.

## III. FIVE INTRINSIC GAPS IN THE CLASSICAL THEORY

Before examining each gap individually, it should be emphasized that the five are not independent. As will become evident, all of them are manifestations of the same root cause: the omission of majority-carrier dynamics and of the QNR electric field from the classical analysis. This mutual interdependence is precisely why a unified treatment is required: resolving any one in isolation inevitably draws in the others.

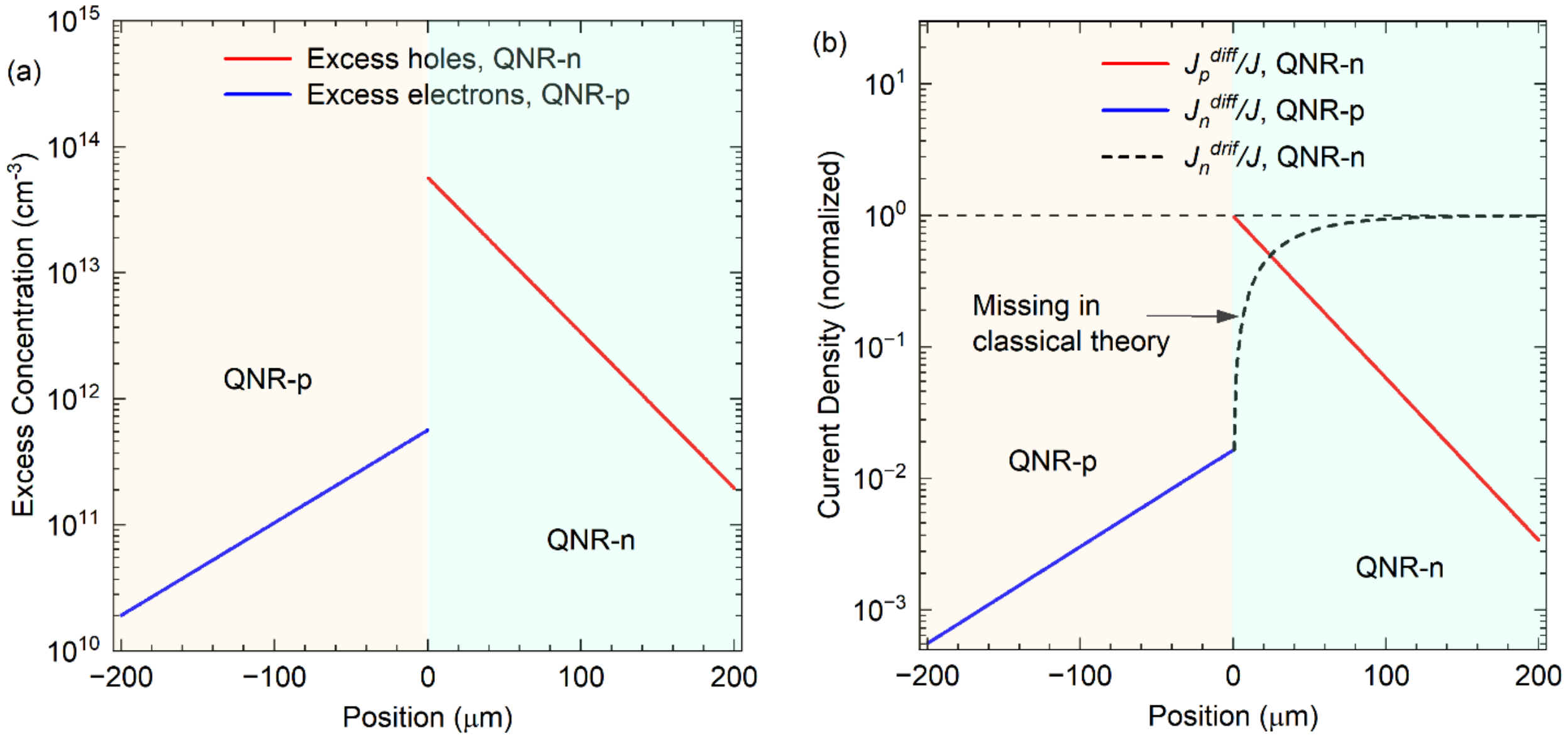

**FIG. 2.** The classical picture of carrier injection under forward bias ($V = 0.5$ V; same device as Fig. 1). (a) Excess-carrier profiles in the two QNRs, each decaying exponentially from the depletion edge with its diffusion length ($L_p = 35\ \mu$m, $L_n = 59\ \mu$m). By charge-quasi-neutrality, the excess majority-carrier profiles coincide with them point by point (indistinguishable because they are almost equal). (b) The minority-carrier diffusion currents, normalized to the terminal current $J$: $J_p^{\mathrm{diff}}(x)$ decays from $0.98J$ at the n-side SCR edge to zero at the contact, and $J_n^{\mathrm{diff}}(x)$ behaves likewise on the p side. The balance $J_n^{\mathrm{drif}}/J \approx 1 - J_p^{\mathrm{diff}}/J$ (gray dashed) is the share of the current the classical theory never accounts for, and is primarily the majority-carrier drift current required to keep the total current (dotted) position-independent. This missing piece motivates Gaps III and IV.

## A. Gap I: The mechanism of charge quasi-neutrality is not explained

The most pervasive assumption in the classical QNR analysis is that of charge quasi-neutrality (CQN): the net charge density in the quasi-neutral base regions is small but nonzero, of order $\varepsilon_s\, dE/dx$, or equivalently, any excess of minority carriers is accompanied by an equal excess of majority carriers. For QNR-n,

$$\Delta n_n(x) \approx \Delta p_n(x). \tag{6}$$

Yet the classical theory does not explain *why* Eq. (6) holds, *how* it is enforced, or *on what timescale* it is established. In standard textbooks the neutrality of the base is presented as a foundational postulate rather than a derived result: it is asserted that injected minority carriers are instantaneously neutralized by an equal number of majority carriers, which redistribute without generating a significant internal field, and this is treated as self-evident rather than derived from the coupled continuity and Poisson equations.

A deeper issue is that Eq. (6) describes a dynamical equilibrium between two carrier populations that are simultaneously injected, diffusing, and recombining. For the equality to hold at every point in steady state, the majority-carrier profile must be enslaved to the minority-carrier profile by some physical coupling mechanism. That mechanism is the QNR electric field: the very field the classical theory sets to zero, acting on the dielectric (Maxwell) relaxation timescale $\tau_d = \varepsilon_s/\sigma$, which is orders of magnitude shorter than the minority-carrier lifetime and is therefore invisible on the recombination timescale. There is thus a logical circularity: the field is zero *because* CQN holds, and CQN holds *because* the field adjusts to enforce it. The classical theory resolves this circularity by assumption rather than by analysis.

## B. Gap II: The electric field in the quasi-neutral region is not determined

The classical theory asserts that $E(x) \approx 0$ in the quasi-neutral regions. Although the quasi-neutral-region field is implicit in the drift–diffusion equations, the classical theory does not determine it, and therefore cannot check whether the field it neglects is in fact consistent with its own assumptions. The omission is not a small correction: without the field, the description cannot be checked for self-consistency. In fact, the field is obtainable from the theory's own equations. Adding Eqs. (1) and (2), using $J = J_n + J_p$ and the CQN condition, gives

$$E(x) = \frac{J - q\left(D_n - D_p\right)\dfrac{d\Delta p_n}{dx}}{q\left(\mu_n n + \mu_p p\right)}. \quad (7)$$

The significance of this field is not trivial. Its existence is required for the majority-carrier current to flow (see Gap III), and it is the mechanism that enforces CQN (Gap I). As Fig. 3 shows for the reference device, the field is small but nonzero: it peaks near the junction, where the diffusion-driven (Dember) component dominates, and decays toward its bulk ohmic value $J/\sigma$; its magnitude is four to six orders below the depletion-region peak field, which explains (but does not justify) the classical neglect. Several textbooks acknowledge in passing that "a small electric field exists in the QNR to drive majority-carrier current" [4, 5, 16] but provide neither a formula nor a calculation.

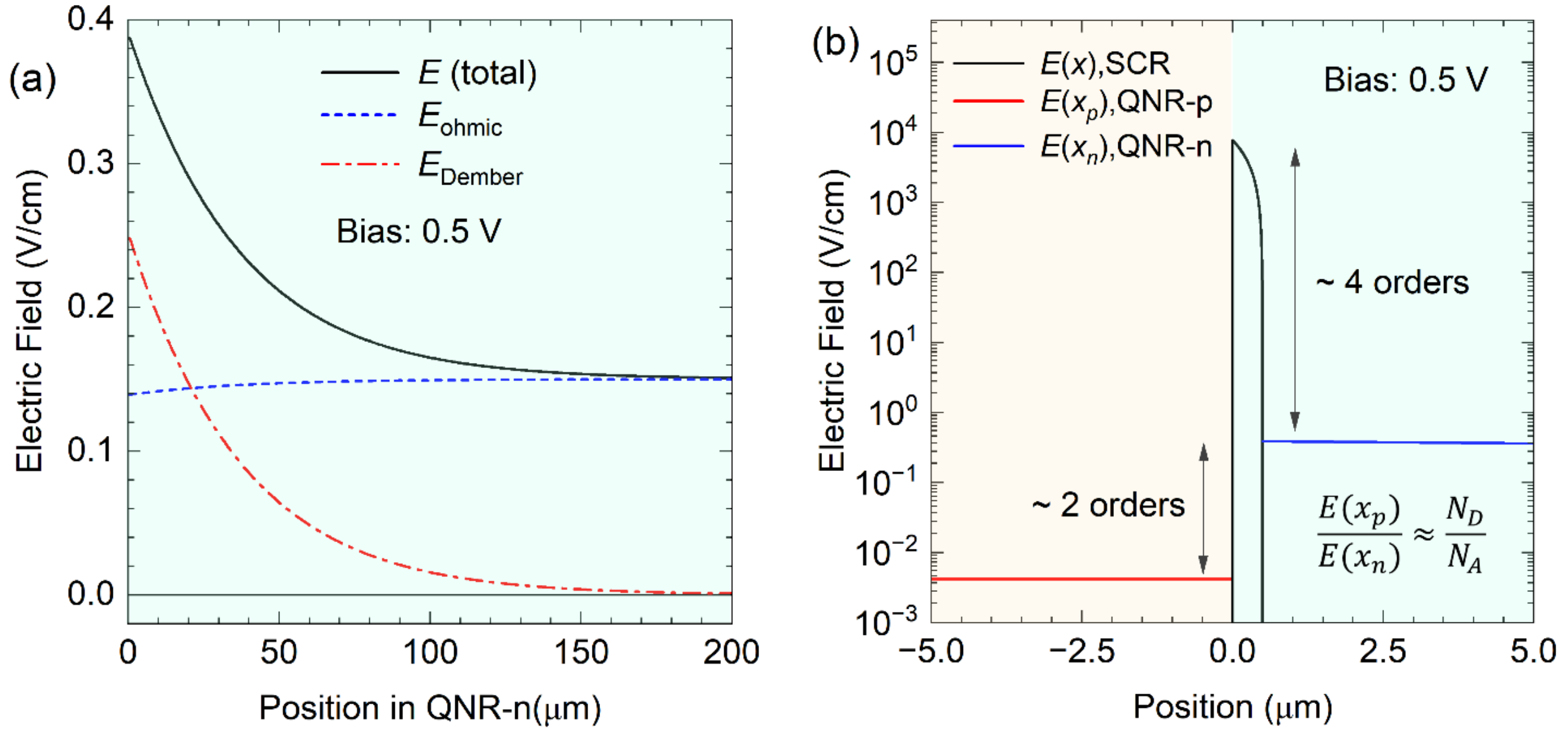

**FIG. 3.** The QNR electric field that the classical theory sets to zero, computed from Eq. (7) for the device of Fig. 1 at $V = 0.5$ V. (a) Spatial profile $E(x)$ in QNR-n: the field is nonzero, peaks near the junction, and decays to its bulk value. (b) Decomposition into the ohmic component $J/\sigma(x)$ (dashed) and the diffusion-driven (Dember) component proportional to $-d\Delta p_n/dx$ (dash-dotted); the Dember part dominates near the junction and vanishes into the bulk. (c) Magnitude comparison on a logarithmic scale across the junction: the QNR fields ($\sim 4 \times 10^{-3}$ V/cm in QNR-p; 0.15-0.4 V/cm in QNR-n) lie four to six orders of magnitude below the depletion-approximation peak field ($\approx 7.7 \times 10^3$ V/cm). Small as it is, the QNR field is what drives the majority-carrier current and enforces charge quasi-neutrality.

This acknowledgment-without-quantification exemplifies how the gap has been recognized but not resolved.

## C. Gap III: The majority-carrier transport is not treated

The classical theory is, at its core, a minority-carrier theory: the electron current in QNR-n and the hole current in QNR-p are never computed. This omission leaves the majority-carrier transport mechanism unspecified and produces two coupled outcomes.

On the one hand, majority-carrier diffusion is missing. In the conventional framework the QNR is treated as field-free ($E = 0$) and transport is reduced to minority-carrier diffusion. But by the CQN condition the excess majority-electron density in QNR-n equals the excess minority-hole density at every point, so the total electron concentration is

$$n_n(x) = n_{n0} + \Delta n_n(x) \approx N_{\mathrm{D}} + \Delta p_n(x). \tag{8}$$

Since $\Delta p_n(x)$ varies with position, Eq. (8) shows that the majority-electron concentration necessarily supports a majority-carrier diffusion current, directed *opposite* to the minority-hole diffusion current because the two profiles share the same gradient. Omitting it obscures how the total current can remain consistent across the device.

On the other hand, majority-carrier drift is missing. The full majority-carrier current density in QNR-n is

$$J_n^{\text{QNR-n}}(x) = q\mu_n n_n(x)\, E(x) + qD_n \frac{d\Delta p_n}{dx}, \tag{9}$$

and by setting $E = 0$ the classical treatment removes the drift term altogether.

Taken together, these omissions imply that the current flowing through the device would be carried only by minority-carrier diffusion, which decays continuously with distance and, in the long-base limit, falls to zero at the ohmic contact. This plainly contradicts physical reality, since the terminal current must be the same at every cross-section of the device. Only a nonzero field $E(x)$ can make the drift term in Eq. (9) act against the majority-carrier diffusion tendency and drive the majority current toward the contact, thereby carrying the share of the total current that the decaying minority-carrier current relinquishes. The classical theory therefore leaves unclear *whether*, *how*, and *to what extent* majority carriers continue the minority-carrier current while maintaining current continuity. In other words, the classical theory leaves unclear the role of the majority carriers, and hence that of the doping.

## D. Gap IV: The current continuity is not rigorously demonstrated

As noted in Sec. II, the governing equations require the total current to be spatially uniform in steady state,

$$J = J_n(x) + J_p(x) = \text{const}, \tag{10}$$

which follows from charge conservation and is not an additional assumption. Yet the classical treatment never demonstrates it explicitly. Because that theory computes only the minority-carrier currents (each of which decays with position), it exhibits currents that visibly vary in space while offering no account of the compensating majority-carrier currents that must keep their sum constant (cf. Fig. 2(b)). The spatial constancy of $J$ is thus asserted on general grounds but never exhibited through the actual carrier currents. The complementary variation of $J_n(x)$ and $J_p(x)$ is the physical content of current continuity, but this is exactly the picture that the classical theory leaves undescribed.

## E. Gap V: The terminal current is not derived from first principles

In the classical theory, the expression for the total current, the diode equation, is in essence a result of logical reasoning rather than a rigorous derivation. The reasoning runs as follows: the minority-carrier diffusion current at either SCR edge is formed by carriers injected from the opposite side; since the carriers from both sides must traverse the SCR to be injected, the two edge currents must add up to the total current in the device:

$$J = J_p^{\text{diff}}(x_{n0}) + J_n^{\text{diff}}(-x_{p0}) = J_0\left(e^{V/V_T} - 1\right), \tag{11}$$

as long as the generation-recombination within the depletion region is negligible.

This is the Shockley diode equation; it is not a derivation from first principles. The rigorous procedure is instead to compute $J = J_n(x) + J_p(x)$ at an *arbitrary* position using the complete expressions for both carrier currents, and to show that the result is independent of $x$ and equals the right-hand side of Eq. (11). Only by such a calculation does the obtained result satisfy the definition of an electric current. To the present author's knowledge, no existing treatment carries out this approach in full generality. This matters in practice: in short-base transistors, heterojunction diodes, and solar cells, the choice of evaluation point affects the result once approximations are introduced, so a derivation valid at any cross-section is precisely what a complete theory should provide.

# IV. SURVEY OF THE LITERATURE

Having identified the five gaps, we now assess the extent to which each has been addressed in the literature. The survey proceeds from the foundational works, through the explicit recognition of the field issue, to textbook and research treatments.

## A. Foundational and early works

The foundational contributions to p-n junction theory were made in a brief and extraordinary period from the late 1940s to the early 1960s. Shockley's 1949 paper [1] remains the essential reference; the extension by Sah, Noyce, and Shockley [6], together with the recombination statistics of Hall [7] and Shockley and Read [8], explains the observed range of ideality factors, and Moll [9] and Ebers and Moll [10] developed the charge-control formalism underlying transistor analysis. None of these pioneering contributions, however, addressed majority-carrier transport or the quasi-neutral-region field in detail. Their focus was minority-carrier injection and the terminal characteristics, which the classical framework handles correctly, so the internal incompleteness was never perceived as a gap, because the diode equation could be obtained without confronting it.

## B. Recognition of the base electric field

The earliest explicit recognition that a field must exist in the QNR came from Fletcher [17], who noted that majority-carrier drift must balance the diffusion tendency and estimated the field magnitude in the high-injection context; his analysis was qualitative and yielded no closed-form low-injection expression. Kroemer [18], in his treatment of drift and graded-base transistors, showed that the coupling between minority and majority carriers through the neutrality condition produces an effective field that modifies the apparent minority-carrier diffusivity, closely related to the concept of ambipolar diffusion [19], introduced to semiconductor physics by van Roosbroeck within his general drift-diffusion formulation [20]. In ambipolar transport the electron-hole pair moves as a unit with effective diffusivity

$$D_a = \frac{(n+p)\,D_n D_p}{n D_n + p D_p}, \tag{12}$$

which in the low-injection limit ($\Delta p_n \ll N_\mathrm{D}$ in QNR-n) reduces to $D_a \approx D_p$, showing why the classical minority-carrier diffusion equation is a good approximation for the minority density. Crucially, the ambipolar framework *also* yields the accompanying field and the majority-carrier current: the very pieces the classical theory omits. Lanyon and Tuft [21] further underscored, in the context of band-gap narrowing and degeneracy at high doping, the need for a complete two-carrier treatment.

## C. Textbook treatments and their limitations

The major textbooks span a spectrum of partial treatments. Sze [2] and Sze and Ng [3] give comprehensive discussions of injection, tunneling, and breakdown but retain the classical minority-carrier framework and mention the QNR field only in connection with heavy doping. Pierret [5] acknowledges in a footnote that "a small electric field exists in the quasi-neutral regions to drive the majority carriers" without quantifying it; this is perhaps the clearest textbook signal that the shortcoming has been noticed but not resolved. Streetman and Banerjee [4] note qualitatively that majority carriers must flow and that the field is small, but provide no analytical treatment. Gray *et al.* [16] and Muller, Kamins, and Chan [22] offer physically instructive discussions of the complementary variation of the two carrier currents and of rapid majority-carrier redistribution, yet neither computes the field nor derives the separate current components. McKelvey [23] and Hauser [24] treat charge neutrality and the boundary conditions with particular care: Hauser's analysis [24] shows the Shockley boundary conditions to be consistent with a more general treatment, but stops short of computing the field. Fonstad [25] correctly identifies that $J_n$ and $J_p$ individually vary with position in the QNR but presents no closed-form expressions. Sah [26] and Neudeck [27] present the classical treatment with exemplary care and completeness, and Pulfrey's monograph [28] gives a modern transistor-oriented account; none of these, however, computes the QNR field or the separate majority-carrier current components.

## D. Research literature contributions

In the research literature, several works have made progress on subsets of the problem. Marshak and van Vliet [29, 30] developed a rigorous thermodynamic formulation of transport in materials with position-dependent band structure, establishing the proper driving forces for current flow; their application was to graded and heterogeneous devices rather than to uniformly doped junctions under low injection. Lundstrom and Schuelke [31] examined minority- and majority-carrier transport in the transistor base with attention to the self-consistent field, confirming numerically the existence and significance of the quasi-neutral base field and providing expressions for the separate carrier currents: this is the closest antecedent among the transistor-oriented treatments, though not framed as a reassessment of the basic junction. Pulfrey and Sanghera [32] extended such analysis to heterojunction bipolar transistors, and De Mari [33] obtained exact numerical solutions of the full drift-diffusion system, confirming that the QNR field is small in low injection yet physically necessary; however, being numerical, and descending from the self-consistent iterative tradition of Gummel [34], such work cannot replace a closed-form analytical framework. Brennan [35] and Neamen [36] treated aspects of the problem pedagogically without resolving the five gaps, and more recent computational and transport-theoretic studies [37–39] have continued to treat the internal field and current-component distributions as numerical outputs rather than analytical results.

In our previous work [40], we reconstructed the classical p-n junction theory by explicitly incorporating excess majority carriers (EMaCs) alongside the conventionally treated excess minority carriers (EMiCs). For a long-base p-n junction we showed analytically that EMaCs and EMiCs are equal at every point in the quasi-neutral base, which is consistent with the CQN condition, and that the EMaCs give rise to an "accompanying electric field" that maintains space-charge neutrality across the junction. That framework yields closed-form expressions for the base field and for all four current components ($J_n^{\mathrm{drift}}, J_n^{\mathrm{diff}}, J_p^{\mathrm{drift}}, J_p^{\mathrm{diff}}$). It recovers a diode equation consistent with Shockley's result, although the minority-carrier drift current was neglected in summing the net current, and the treatment was restricted to long-base junctions at low-level injection. That study was the first to treat EMaCs as a primary quantity in junction analysis, and it provides the starting point for the closed-form treatment developed here.

## E. Assessment

The summary of this survey is unambiguous: while individual gaps have been partially recognized, and some partially addressed in specialized contexts, no existing work provides a complete, self-consistent, analytically closed treatment that simultaneously resolves all five gaps within the standard low-injection, uniform-doping framework, retaining all four current components and requiring no auxiliary approximations beyond those of the classical theory itself. Table I summarizes this assessment.

**TABLE I.** Assessment of representative works with respect to the five intrinsic gaps of the classical p–n junction theory.

| Work [Reference] | Gap I: CQN Mechanism | Gap II: QNR Field | Gap III: Majority Transport | Gap IV: Current Continuity | Gap V: Terminal Current |
|---|---|---|---|---|---|
| Shockley (1949) [1] | △ | △ | △ | △ | △ |
| van Roosbroeck (1953) [19] | △ | ✓[a] | ✓ | △ | — |
| Kroemer (1954) [18] | △ | △ | △ | ✗ | ✗ |
| Fletcher (1957) [17] | △ | △ | △ | ✗ | ✗ |
| Gray et al. (1964) [16] | △ | △ | △ | △ | ✗ |
| De Mari (1968) [33] | △ | △ | △ | △ | △ |
| Hauser (1971) [24] | △ | ✗ | ✗ | ✗ | △ |
| Lundstrom & Schuelke (1983) [31][b] | △ | △ | △ | △ | △ |
| Pierret (1996) [5] | △ | △ | △ | △ | ✗ |
| Sze & Ng (2007) [3] | △ | △ | △ | △ | ✗ |
| Streetman & Banerjee (2015) [4] | △ | △ | △ | △ | ✗ |
| Yang et al. (2017) [40] | ✓ | ✓ | ✓ | △ | △ |

Description: ✓ fully resolves or directly addresses the gap with closed-form analytical expressions; △ partially addresses the gap, provides qualitative discussion, or recognizes the issue without fully resolving it; ✗ does not treat the gap or mentions it only in passing without substantive treatment; — not applicable or not relevant to that gap. CQN: charge quasi-neutrality; QNR: quasi-neutral region. No existing work resolves all five gaps simultaneously in closed analytical form. [a] General closed-form expression for homogeneous material under charge neutrality (ambipolar transport theory). [b] Minority- and majority-carrier transport in a heterojunction transistor base.

# V. TOWARD A SELF-CONSISTENT TWO-CARRIER ANALYTICAL FRAMEWORK

## A. The closure requirements

The five gaps identified above define a single closure problem rather than five independent corrections. Charge quasi-neutrality constrains the relation between the carrier populations; the base electric field couples their transport; and the resulting current partition connects local charge conservation to the terminal response. A self-consistent analytical description must therefore determine carrier densities, electric field, and both carrier currents within the same boundary-value problem, using the standard Poisson, drift–diffusion, and continuity equations. The required extension concerns analytical closure, not new transport physics.

Electrostatic consistency requires distinguishing quasi-neutrality from exact neutrality. A small residual space charge can support a spatially varying field even when the excess electron and hole densities are equal to leading order. Accordingly, quasi-neutrality must be justified as a controlled approximation, and the base field must be determined consistently to the retained order rather than set to zero a priori. Its significance must be assessed separately for each carrier species: a field negligible for minority-carrier transport can remain essential for majority-carrier conduction. This distinction links the first two gaps directly to the third.

Transport consistency requires retaining the drift and diffusion contributions of both carrier species until their relative magnitudes have been established. Their spatial redistribution must satisfy the two continuity equations and explicitly yield a position-independent total current in steady state. Current conservation is already a consequence of those equations; the additional analytical requirement is to exhibit how the resolved current components satisfy it. The same solution must connect this local current balance to the terminal current through consistent junction-edge and contact conditions, thereby addressing the fourth and fifth gaps without treating terminal-current evaluation as a separate construction. The device relevance of these requirements is illustrated by carrier injection and recombination, conductivity modulation, and internal voltage drops in junction-based bipolar power devices [41] and optoelectronic devices [42]. These examples motivate a consistent connection between internal transport and terminal characteristics, rather than an extension of the present analysis to those devices.

Such a framework should recover the classical diode equation under its established assumptions while specifying when the neglected field or current contributions become significant. Its value lies not in replacing the successful classical limit, but in making the connection between electrostatics, two-carrier transport, and terminal response explicit within a stated domain of validity. This provides a unified physical criterion for closing the five gaps without presupposing a particular closed-form solution. Fig. 4 summarizes these interdependent requirements as a framework for analytical closure, rather than a sequence of independent corrections.

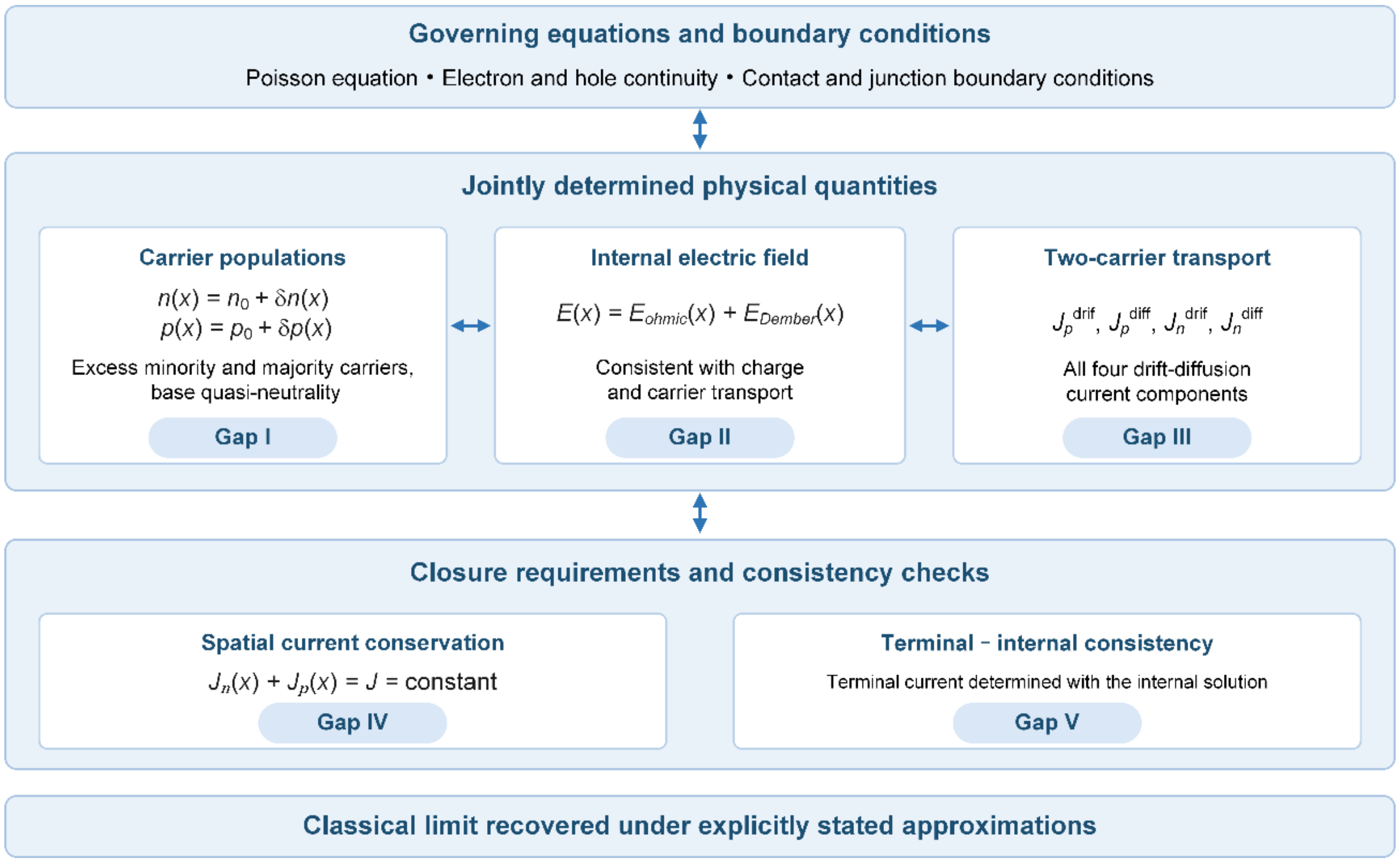

**FIG. 4.** Conceptual framework for a self-consistent two-carrier analytical description of a p–n junction. Carrier populations, the internal electric field, and electron and hole drift–diffusion must be determined jointly under the governing equations and device boundary conditions. These coupled quantities address Gaps I–III, while spatial current conservation and consistency between the internal solution and terminal current provide the closure requirements associated with Gaps IV and V. Recovery of the classical limit requires explicitly stated approximations. The diagram specifies requirements for analytical closure rather than results established in this paper.

## B. What the requirements imply

Stated as requirements, the five gaps are easy to accept and hard to distinguish from a restatement of the governing equations. Their content becomes visible only when the requirements are imposed together and the resulting description is examined as a whole. Three structural features then follow from the equations themselves, without an additional hypothesis, and each of them replaces an assumption of the classical derivation.

The first concerns the junction edges. Applied to both carrier species, the law of mass action together with quasi-neutrality returns a single relation between the boundary excess and the junction voltage. The two edges are therefore not independent: they are fixed by one variable, and the two boundary populations are locked to each other. The classical treatment reaches the same place by assigning a boundary condition separately at each edge and then identifying the two junction voltages by assumption.

The second concerns the field. Requiring the sum of the four current components to be divergence-free in steady state fixes the internal field as the sum of two terms, a drift term set by the local conductivity and the total current, and a diffusion term set by the imbalance between the electron and hole diffusivities. The magnitude is small, some tenths of a volt per centimetre for a silicon junction under forward bias and four to six decades below the peak field in the space-charge region, but it is precisely the magnitude required to carry the majority current that the minority diffusion current sheds along the base.

The third concerns the current. Once the profile and the field are known, all four components follow in closed form and their position-dependent parts cancel identically. The total current is then uniform across the device as a consequence of the solution rather than as an appeal to charge conservation, and the terminal current is one reading of a quantity already known to be independent of position.

These features move four quantities from assumptions to results of the boundary-value problem: quasi-neutrality follows from electrostatics to the retained order, the base field is solved for rather than set to zero, majority transport appears explicitly, and total current is uniform by construction. The terminal current can then be read at any cross-section of that solution.

## C. Outlook

Three directions follow from this structure. For power devices, treating the applied bias as a sum of a junction drop and a base drop would allow the dissipation profile to be located within the structure rather than assigned to a fitted resistance. For compact modelling, a voltage partition computed from doping, geometry and lifetime could replace a fitted series resistance, and the ideality factor would follow from the same solution rather than from a separate empirical rule. For the theory itself, graded and heterojunction structures, transient conditions, and systems of lower carrier density each require the boundary relation to be rebuilt, while the closure structure remains unchanged. The value of stating the requirements in this form is not that these problems are solved, but that they supply a common checklist: any candidate treatment can be asked whether it determines the carrier populations, the internal field, both carrier currents, their spatial balance and the terminal current, or whether it still assumes one of them.

# VI. CONCLUSION

This paper has examined five interconnected gaps in the conventional minority-carrier-based analytical description of p–n junctions. They concern the justification of quasi-neutrality, determination of the base electric field, resolution of majority-carrier transport, explicit spatial balance of the carrier currents, and connection of that balance to terminal current. Their common origin is an incomplete analytical coupling of electrostatics and two-carrier transport, rather than missing physical laws. The contribution of this analysis is to organize these issues into a single closure problem and identify the physical requirements for its resolution.

Two distinctions are essential. First, quasi-neutrality does not imply a vanishing electric field: a small residual space charge can support a spatially varying field, and a field negligible for minority-carrier transport can remain essential for majority-carrier conduction. Second, steady-state current conservation follows from the continuity equations, whereas a complete analytical description must additionally exhibit how the spatially varying drift and diffusion components satisfy it. Likewise, the conventional junction-edge evaluation of terminal current is justified under the ideal-diode assumptions; the unresolved task is to connect that result explicitly to the local two-carrier current throughout the device.

A self-consistent analytical framework should therefore determine carrier populations, electric field, and both carrier currents within one boundary-value problem, with approximations controlled to a consistent order. It should recover the classical limit and state where that reduction ceases to be adequate. This provides a device-physics framework for further theoretical development.

# ACKNOWLEDGMENTS

The authors thank Dr. Guipeng Liu (Lanzhou University), for helpful suggestions.

## DATA AVAILABILITY

There are no publicly available research data or software supporting this manuscript. Requests for further information or data should be sent to the authors.

---

[1] W. Shockley, The theory of p-n junctions in semiconductors and p-n junction transistors, *Bell Syst. Tech. J.* **28**, 435 (1949).

[2] S. M. Sze, *Physics of Semiconductor Devices*, 2nd ed. (Wiley-Interscience, New York, 1981).

[3] S. M. Sze and K. K. Ng, *Physics of Semiconductor Devices*, 3rd ed. (Wiley, Hoboken, NJ, 2007).

[4] B. G. Streetman and S. K. Banerjee, *Solid State Electronic Devices*, 7th ed. (Pearson, Upper Saddle River, NJ, 2015).

[5] R. F. Pierret, *Semiconductor Device Fundamentals* (Addison-Wesley, Reading, MA, 1996).

[6] C. T. Sah, R. N. Noyce, and W. Shockley, Carrier generation and recombination in p-n junctions and p-n junction characteristics, *Proc. IRE* **45**, 1228 (1957).

[7] R. N. Hall, Electron-hole recombination in germanium, *Phys. Rev.* **87**, 387 (1952).

[8] W. Shockley and W. T. Read, Statistics of the recombinations of holes and electrons, *Phys. Rev.* **87**, 835 (1952).

[9] J. L. Moll, *Physics of Semiconductors* (McGraw-Hill, New York, 1964).

[10] J. J. Ebers and J. L. Moll, Large-signal behavior of junction transistors, *Proc. IRE* **42**, 1761 (1954).

[11] J. A. del Alamo and R. M. Swanson, Forward-bias tunneling: A limitation to bipolar device scaling, *IEEE Electron Device Lett.* **7**, 629 (1986).

[12] R. People, Indirect band gap of coherently strained $Ge_xSi_{1-x}$ bulk alloys on ⟨001⟩ silicon substrates, *Phys. Rev. B* **32**, 1405 (1985).

[13] A. Neugroschel, C. T. Sah, and M. S. Carroll, Diffusion length and lifetime measurements in silicon p-n junction solar cells, *J. Appl. Phys.* **46**, 5325 (1975).

[14] J. G. Fossum, Computer-aided numerical analysis of silicon solar cells, *Solid-State Electron.* **19**, 269 (1976).

[15] W. Shockley, *Electrons and Holes in Semiconductors* (Van Nostrand, New York, 1950).

[16] P. E. Gray, D. DeWitt, A. R. Boothroyd, and J. F. Gibbons, *Physical Electronics and Circuit Models of Transistors* (Wiley, New York, 1964).

[17] N. H. Fletcher, The high current limit for semiconductor junction devices, *Proc. IRE* **45**, 862 (1957).

[18] H. Kroemer, Zur Theorie des Diffusions- und des Drifttransistors, *Arch. Elektr. Übertrag.* **8**, 223 (1954).

[19] W. van Roosbroeck, The transport of added current carriers in a homogeneous semiconductor, *Phys. Rev.* **91**, 282 (1953).

[20] W. van Roosbroeck, Theory of the flow of electrons and holes in germanium and other semiconductors, *Bell Syst. Tech. J.* **29**, 560 (1950).

[21] H. P. D. Lanyon and R. A. Tuft, Bandgap narrowing in moderately to heavily doped silicon, *IEEE Trans. Electron Devices* **26**, 1014 (1979).

[22] R. S. Muller, T. I. Kamins, and M. Chan, *Device Electronics for Integrated Circuits*, 3rd ed. (Wiley, New York, 2003).

[23] J. P. McKelvey, *Solid State and Semiconductor Physics* (Harper & Row, New York, 1966).

[24] J. R. Hauser, Boundary conditions at p-n junctions, *Solid-State Electron.* **14**, 133 (1971).

[25] C. G. Fonstad, *Microelectronic Devices and Circuits*, 2006 ed. (McGraw-Hill, New York, 1994).

[26] C. T. Sah, *Fundamentals of Solid-State Electronics* (World Scientific, Singapore, 1991).

[27] G. W. Neudeck, *The PN Junction Diode*, Vol. 2 of the Modular Series on Solid State Devices (Addison-Wesley, Reading, MA, 1989).

[28] D. L. Pulfrey, *Understanding Modern Transistors and Diodes* (Cambridge Univ. Press, Cambridge, 2010).

[29] A. H. Marshak and K. M. van Vliet, Electrical current in solids with position-dependent band structure, *Solid-State Electron.* **21**, 417 (1978).

[30] A. H. Marshak and K. M. van Vliet, Electrical current and carrier density in degenerate materials with nonuniform band structure, *Proc. IEEE* **72**, 148 (1984).

[31] M. S. Lundstrom and R. J. Schuelke, Numerical analysis of heterostructure semiconductor devices, *IEEE Trans. Electron Devices* **30**, 1151 (1983).

[32] D. L. Pulfrey and H. S. Sanghera, Minority and majority carrier transport in heterojunction bipolar transistors, *Solid-State Electron.* **36**, 495 (1993).

[33] A. De Mari, An accurate numerical steady-state one-dimensional solution of the p-n junction, *Solid-State Electron.* **11**, 33 (1968).

[34] H. K. Gummel, A self-consistent iterative scheme for one-dimensional steady state transistor calculations, *IEEE Trans. Electron Devices* **11**, 455 (1964).

[35] K. F. Brennan, *The Physics of Semiconductors: With Applications to Optoelectronic Devices* (Cambridge Univ. Press, Cambridge, 1999).

[36] D. A. Neamen, *Semiconductor Physics and Devices: Basic Principles*, 4th ed. (McGraw-Hill, New York, 2012).

[37] M. Lundstrom and J. Guo, *Nanoscale Transistors: Device Physics, Modeling and Simulation* (Springer, New York, 2006).

[38] S. Selberherr, *Analysis and Simulation of Semiconductor Devices* (Springer, Vienna, 1984).

[39] M. S. Lundstrom, *Fundamentals of Carrier Transport*, 2nd ed. (Cambridge Univ. Press, Cambridge, 2000).

[40] J. Yang, Y. Zhang, W. Chen, and G. Liu, p-n junction theory in view of excess majority carriers, *EPL* **120**, 28004 (2017).

[41] B. J. Baliga, *Fundamentals of Power Semiconductor Devices* (Springer, New York, 2008).

[42] A. Luque and S. Hegedus, Eds., *Handbook of Photovoltaic Science and Engineering*, 2nd ed. (Wiley, Chichester, 2011).